\documentstyle[twocolumn,aps,psfig]{revtex}

\begin{document}
\title{A precise approximation for directed percolation in $d=1+1$}
\author{Cl\'ement Sire} 
\address{Laboratoire de Physique Quantique (UMR 5626 du CNRS)\\
Universit\'e Paul Sabatier, F-31062 Toulouse Cedex 4, France\\
Clement.Sire@irsamc.ups-tlse.fr}

\maketitle
\begin{center}
{\bf Abstract}
\end{center}
\vskip -0.3cm
\begin{abstract}
We introduce an approximation specific to a continuous model for
directed percolation, which is strictly equivalent to $1+1$
dimensional directed bond percolation. We find that the critical
exponent associated to the order parameter (percolation probability)
is $\beta=\frac{1}{2}\left(1-\frac{1}{\sqrt{5}}\right)=
0.276393202...$, in remarkable agreement with the best current
numerical estimate $\beta=0.276486(8)$.
\end{abstract}

%\pacs{05.70.Ln, 64.60.Ht, 64.60.Ak}
\vskip 0.5cm 
\narrowtext
Directed percolation (DP) \cite{dp1,dp2,dpreview,dpapp} is a useful
paradigm for dynamical phase transitions between an active/spreading
phase and an extinct/absorbing phase. Models in the DP class of
universality are involved in the description of catalytic reactions
\cite{cata}, surface dynamics \cite{surf}, porous systems
\cite{dpapp}, granular media \cite{sand}, epidemics, 
Calcium dynamics in cells \cite{dpapp}, developed
turbulence and coupled maps \cite{dppomeau}...  Recently,
H. Hinrichsen summarized the large scope of possible physical
applications of DP \cite{dpapp}, which led P. Grassberger to
conjecture that the DP universality class should describe any
continuous phase transition from a fluctuating active phase into a
{\it single} absorbing phase, in the absence of quenched disorder and
special symmetries \cite{gen}. In a sense, DP plays a similar role in
the study of dynamical phase transitions as the Ising model for
continuous equilibrium phase transitions \cite{dpapp}.

Despite its ubiquity, DP is maybe the only major statistical physics
model which has not yet been successfully solved in one spatial
dimension (+ time), probably due to its lack of conformal invariance.

Let us recall the original model of directed bond percolation in
$d=1+1$, describing the propagation of a fluid in a $2d$ porous
medium. On a square lattice tilted at $45^\circ$, a fraction of bonds
$p$ are chosen at random to be active, whereas the remaining bonds
stay inactive or broken. The ``fluid'' starts from the top row, and
propagates downward, only passing through the active bonds. One then
defines the order parameter $n(p,t)$, which measures the average
density of occupied sites at row $t$. $n(p,t)$ happens to coincide
with the probability that at least one site at row $t$ is still active
(percolation probability).

In the stationary limit $t\to+\infty$, the order parameter tends to a
constant value $n(p)$, which is zero below $p_c=0.644700185(5)$
\cite{pc,beta}, and behaves as $n(p)\sim (p-p_c)^\beta$, near
$p_c$. This defines the universal critical exponent
$\beta=0.276486(8)$ \cite{pc,beta}. Defining $n_i(t)=1$
(resp. $n_i(t)=0$), when the $i^{\rm th}$ site at row $t$ is active/occupied (resp. inactive/empty), $n_i(t)$ satisfies the following recursion relation
\begin{equation}
n_i(t+1)=a_i n_i(t)+b_i n_{i+1}(t)-a_i b_i n_i(t)n_{i+1}(t),
\label{ndp}
\end{equation}
where $a_i(t)$ and $b_i(t)$ are independent random variables taking
the value 1 with probability $p$ (if the corresponding ongoing bond is
active), and 0 otherwise.

In relation to self-organized criticality (SOC), it has been
recognized that directed bond percolation is strictly equivalent to a
continuous dynamical model (SOCDP), involving no external parameter
(like $p$ in DP) \cite{cont1,cont2,cont3,cont4}. On a $1d$ lattice, we
define the continuous variables $x_i(t)$ as satisfying the recursion
relation
\begin{equation}
x_i(t+1)=\min\left[\max(x_i(t),z_i),\max(x_{i+1}(t),z_i')\right],
\label{xdp}
\end{equation}
where $z_i(t)$ and $z_i'(t)$ are independent random variables
uniformly distributed between 0 and 1.  It can be easily shown that
the $n_i$'s for directed bond percolation and the $x_i$'s are very
simply related:
\begin{equation}
n_i(t)=\theta(x_i(t)-p),
\end{equation}
where $\theta(.)$ is the usual Heaviside step function.  In the large
time limit, the $x_i$'s are distributed according to a stationary
probability distribution $\rho(x)$, and
\begin{equation}
n(p)=\int_{0}^{p}\rho(x)\,dx.
\end{equation}
Now, using Eq.~(\ref{ndp}) (or equivalently Eq.~(\ref{xdp})), we find
in the stationary limit,
\begin{eqnarray}
\frac{2p-1}{p^2}n(p)&&=\frac{2p-1}{p^2}\langle n_1\rangle=
\langle n_1n_2\rangle,\label{eqcor}\\
&&=\int_0^p\int_0^p\rho_2(x_1,x_2)\,dx_1dx_2,
\end{eqnarray}
where $\rho_2(x_1,x_2)$ is the nearest neighbor correlation function
of the $x_i$'s.  In mean field (MF) theory, one makes the
approximation $\langle n_1n_2\rangle\approx \langle n_1\rangle^2$,
leading to
\begin{equation}
n_{MF}(p)=\frac{2p-1}{p^2},\quad
\rho_{MF}(p)=\frac{2(1-p)}{p^3}.\label{MF}
\end{equation}

From now, we study the stationary state of DP in terms of the
continuous model defined by Eq.~(\ref{xdp}). We first notice that
\begin{eqnarray}
x_1(t+1)=&&\min[x_1(t),x_2(t)], {\rm \ with\ probability\
}\label{pmin}\\ p_{\rm min}=&&\min[x_1(t),x_2(t)],\nonumber\\
x_1(t+1)=&&\max[x_1(t),x_2(t)], {\rm \ with\ probability\
}\label{pmax}\\ p_{\rm
max}=&&\max[x_1(t),x_2(t)](1-\max[x_1(t),x_2(t)]),\nonumber
\end{eqnarray}
so that there is a non zero probability that $x_1(t+1)=x_2(t+1)$
exactly. Hence, the two-point correlation function of the $x_i$'s,
$\rho_2(x_1,x_2)$, should include a $\delta(x_1-x_2)$ contribution
($\delta(.)$ is the Dirac peak distribution). Thus, in all generality,
we write $\rho_2(x_1,x_2)$ in the following form:
\begin{equation}
\rho_2(x_1,x_2)=\tilde{\rho}_2(x_1,x_2)+\rho(x_1)g(x_1)\delta(x_1-x_2),
\end{equation}
which defines $g(p)$ as the probability that $x_2=p$, conditional to
the fact that its neighbor $x_1=p$.  We then define $f(x_1,x_2)$
through the relation
\begin{equation}
\tilde{\rho}_2(x_1,x_2)=\rho(\min(x_1,x_2))f(x_1,x_2),\label{f12} 
\end{equation}
noting that as $\rho(p)$ diverges near $p_c$ (since $\beta<1$),
$\rho(\min(x_1,x_2))>\rho(\max(x_1,x_2))$, at least near $p_c$
(numerically $\rho(p)$ appears to be a strictly decreasing function,
like in mean field theory). We expect that $f(x_1,x_2)$ is a smooth
function of order unity. Indeed, contrary to the MF approach (where
one assumes that $\langle n_1n_2\rangle\sim [n(p)]^2$), correlation
functions all behave as $n(p)$ near $p_c$. Indeed, as $p_c>1/2$,
Eq.~(\ref{eqcor}) implies that $\langle n_1n_2\rangle\sim n(p)\sim
(p-p_c)^\beta$, such that $\tilde{\rho}_2(p,p)\sim \rho(p)\sim
(p-p_c)^{-(1-\beta)}$ (instead of $\rho(p)^2\sim
(p-p_c)^{-2(1-\beta)}$, predicted by MF). A natural guess for
$f(x_1,x_2)$ is provided by the general statement that although MF is
inept at describing correlation functions near $p_c$, it still leads to
reasonably accurate amplitude ratios between them, for all values of
$p$ (at least for short range correlation functions like $\langle
n_1n_2\rangle$). Hence, this prompts the introduction of the key 
approximation
\begin{eqnarray}
f(x_1,x_2)&&\approx f_{MF}(x_1,x_2),\\
&&=\frac{\rho_{MF}(x_1)\rho_{MF}(x_2)}{\rho_{MF}(\min(x_1,x_2))},\\
&&= \rho_{MF}(\max(x_1,x_2)).\label{fcm}
\end{eqnarray}
In the following, we make the more general {\it ansatz}
\begin{equation}
\tilde{\rho}_2(x_1,x_2)=\rho(\min(x_1,x_2))f(\max(x_1,x_2)),
\label{ansatz}
\end{equation}
where $f(p)$ is not necessarily equal to $\rho_{MF}(p)$.
$\rho(p)$, $f(p)$ and $g(p)$ are not independent functions as they are
related together by Eq.~(\ref{eqcor}), and by the probability conservation
constraint, 
\begin{equation}
\rho(p)=\int_0^1\rho_2(x,p)\,dx. \label{prob1}
\end{equation}
From Eq.~(\ref{eqcor}) and Eq.~(\ref{prob1}), and after straightforward
calculations, we obtain the two relations
\begin{eqnarray}
&&n(p)=\exp\left[-\int_p^1\frac{2f(x)-\rho_{MF}(x)}{n_{MF}(x)-g(x)}\,dx
\right],\label{rel1}\\ &&g(p)+\int_p^1f(x)\,dx
+f(p)\frac{n(p)}{\rho(p)}=1.\label{rel2}
\end{eqnarray}
From Eq.~(\ref{rel1}) and Eq.~(\ref{rel2}), one can obtain a first
order differential equation for $F(p)=f(p)-\rho_{MF}(p)$, involving
only $p$ and $g(p)$. This equation can be shown to have only $F\equiv
0$ as a global solution satisfying the boundary conditions and the physical
constraints. Hence, and in complete accordance with the physical argument given in Eq.~(\ref{fcm}), we find
\begin{equation}
f(p)=\rho_{MF}(p).
\end{equation}
Quite remarkably, for this precise form for $f(p)$, Eq.~(\ref{rel2})
is now satisfied for any choice of $g(p)$, so that we are left with
Eq.~(\ref{rel1}) as the only non trivial relation between $n(p)$ and
$g(p)$.

Now, as $n(p)$ vanishes at $p_c$, we expect the function involved in
the integral of Eq.~(\ref{rel1}) to develop a single pole at $p_c$,
of residue $\beta$, so that $n(p)\sim(p-p_c)^\beta$, near $p_c$. This
leads to
\begin{eqnarray}
g(p_c)=&&n_{MF}(p_c)=\frac{2p_c-1}{p_c^2},\label{rel1pc}\\
\beta=&&\left(1-\frac{g'(p_c)}{\rho_{MF}(p_c)}\right)^{-1}.\label{rel2pc}
\end{eqnarray}
Note that Eq.~(\ref{rel1pc}) is in fact an exact identity, which does
not rely on the present approximation on $\tilde{\rho}_2(x,y)$.

In order to achieve our goal of computing $n(p)$, we need a further
relation for $g(p)$. This can be obtained by writing the exact
stationary equation for $g(p)$. We first define
$\langle\bullet\bullet\bullet\rangle$ as the probability of having
three consecutive sites with $x_i=p$, divided by $\rho(p)$ (that is,
conditional to having one site with $x_i=p$). In the same manner, we define
$\langle\bullet\bullet\overline{p}\rangle$
(resp. $\langle\bullet\bullet\underline{p}\rangle$) as the probability
of having two consecutive sites with $x_1=x_2=p$, and $x_3>p$
(resp. $x_3<p$), divided by $\rho(p)$. Finally, in the stationary
limit, we obtain the exact relation
\begin{eqnarray}
g(p)=&&(2p-p^2)^2\langle\bullet\bullet\bullet\rangle\label{gpt}\\
&&+2(2p-p^2)\left(p\langle\bullet\bullet\overline{p}\rangle
+p(1-p)\langle\bullet\bullet\underline{p}\rangle\right)\nonumber\\
&&+p^2\langle\overline{p}\bullet\overline{p}\rangle
+2p^2(1-p)\langle\underline{p}\bullet\overline{p}\rangle
+p^2(1-p)^2\langle\underline{p}\bullet\underline{p}\rangle\nonumber\\
&&+p^2\langle\bullet\,\overline{p}\,\bullet\rangle\nonumber\\
=&&\langle\bullet\,\bullet\rangle.\nonumber
\end{eqnarray}
For instance, the first term
$(2p-p^2)^2\langle\bullet\bullet\bullet\rangle$ represents the fact
that a configuration $\bullet\,\bullet$ $(x_1(t+1)=x_2(t+1)=p)$ at
time $t+1$ can arise from a configuration $\bullet\bullet\bullet$
$(x_1(t)=x_2(t)=x_3(t)=p)$ at time $t$, provided that $x_1$ and $x_2$
are preserved by the transformation of Eq.~(\ref{xdp}). This happens
with probability
\begin{equation}
(p_{\rm min}+p_{\rm max})^2=(p+p(1-p))^2=(2p-p^2)^2,
\end{equation}
hence the coefficient in Eq.~(\ref{gpt}) ($p_{\rm min}$ and $p_{\rm
max}$ have been defined in Eq.~(\ref{pmin}) and Eq.~(\ref{pmax})).

Eq.~(\ref{gpt}) relates $g(p)$ to three-point correlation functions,
and cannot be exploited unless an additional approximation is
introduced. We will factor these three-point correlation functions
into products of two-point correlation functions, according to the
usual mean field scheme. Introducing $p_+$ as the probability that
$x_2>p$, conditional to the fact that $x_1=p$, we obtain
\begin{eqnarray}
[1-g(p)]p_+&&=\frac{\int_p^1 \tilde{\rho}_{2}(x,p)\,dx}{\rho(p)},\label{pp}\\
&&=\int_p^1 \rho_{MF}(x)\,dx=\frac{(1-p)^2}{p^2},
\end{eqnarray}
where Eq.~(\ref{pp}) is an exact identity.  We give below a few
examples of three-point correlation functions computed according to
this MF factorization scheme:
\begin{eqnarray}
&&\langle\bullet\bullet\bullet\rangle=g(p)^2,\label{gp2}\\
&&\langle\bullet\bullet\overline{p}\rangle=g(p)[1-g(p)]p_+=
g(p)\frac{(1-p)^2}{p^2},\label{gpp}\\
&&\langle\underline{p}\bullet\overline{p}\rangle=[1-g(p)]^2p_+(1-p_+),\\
&&\hspace{1.05cm}=\left[\frac{2p-1}{p^2}-g(p)\right]\frac{(1-p)^2}{p^2}.
\end{eqnarray}
Inserting the MF form for the three-point correlation functions into
Eq.~(\ref{gpt}), we finally obtain a closed equation for $g(p)$,
\begin{equation}
g(p)=(1-p+pg(p))^2+(1-p)^2g(p)(1-g(p)),
\end{equation}
which can be readily solved, leading to
\begin{equation}
g(p)=\frac{(1-p)^2}{2p-1}.
\end{equation}

\begin{figure}[hhh]
\psfig{file=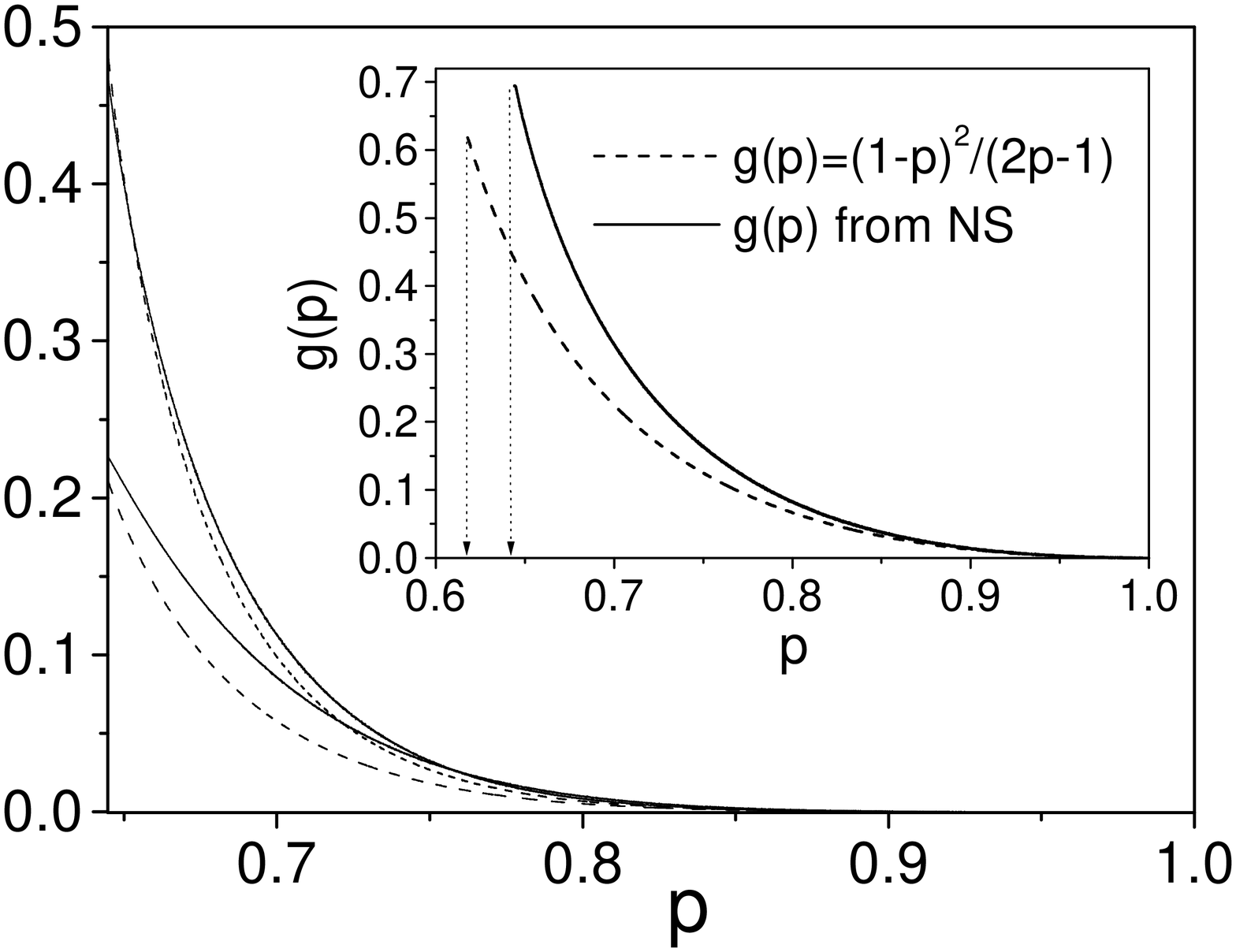,width=9cm}
\caption[]{We plot $\langle\bullet\bullet\bullet\rangle$ 
(two top lines) and $\langle\bullet\bullet\overline{p}\rangle$ as
obtained from numerical simulations (full lines), and as given by
Eq.~(\ref{gp2}) and Eq.~(\ref{gpp}) (dashed lines), where the
numerical value of $g(p)$ has been inserted in these expressions. Note
that $\langle\bullet\bullet\bullet\rangle \approx g(p)^2$, especially
near $p_c$. These functions all vanish as $(1-p)^4$ near $p=1$, as
predicted by Eq.~(\ref{gp2}) and Eq.~(\ref{gpp}). Insert: comparison
between the numerical $g(p)$ and the present theory. In all figures of
this letter, we have simulated a system of $N=300000$ sites, averaged
over 100 samples. Physical quantities in the stationary state have
been estimated by averaging them  between $t=300000$ and $t=310000$.}
\end{figure} 

$p_c$ and $\beta$ can now be calculated by expressing the conditions of
Eq.~(\ref{rel1pc}) and Eq.~(\ref{rel2pc}). We obtain
\begin{eqnarray}
p_c&=&g(p_c)=\tau=\frac{\sqrt{5}-1}{2}=0.618033989...,\\
\beta&=&\frac{1}{2}\left(1-\frac{1}{\sqrt{5}}\right)=
0.276393202...,
\end{eqnarray}
where $\tau$ is the golden mean. $p_c$ is only in fair agreement with
the best numerical estimate $p_c=0.644700185(5) $\cite{pc,beta}, although
this represents a dramatic improvement when compared to the mean field
value $p_c=1/2$. Note that the exact identity Eq.~(\ref{rel1pc})
implies that getting $p_c>1/2$ necessitates the introduction of a non
trivial function $g(p)$, which is zero in MF.  The $\beta$ exponent is
in extraordinary agreement (relative accuracy of 0.034\%) with the
best available numerical value $\beta=0.276486(8)$ \cite{pc,beta}.

\begin{figure}[hhh]
\psfig{file=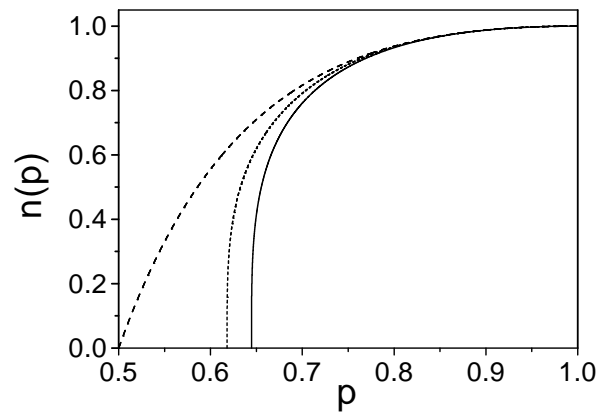,width=9cm}       
\caption[]{We respectively plot $n(p)$ as given 
by mean field theory (dashed line), the present theory (dotted line), and
numerical simulations (full line). Note that the three curves coincide
near $p=1$, as mean field theory becomes exact in this limit. A fit of the
numerical data to the functional form of Eq.~(\ref{nth}), where $\tau$
becomes a fitting parameter, cannot be distinguished from the actual
data.}
\end{figure} 

The fact that the relation
$\langle\bullet\bullet\bullet\rangle=g(p)^2$ seems to be exactly
satisfied numerically at $p_c$ could explain this agreement, which
also implies that the MF factorization of Eq.~(\ref{gpt}) is
quantitatively correct near $p_c$ (the three-point correlation
functions appearing in Eq.~(\ref{gpt}) are not independent and are
related to $\langle\bullet\bullet\bullet\rangle$ by various sum
rules).  This is illustrated in Fig.~1, where the exact numerical
$g(p)=\langle\bullet\,\bullet\rangle$,
$\langle\bullet\bullet\bullet\rangle$, and
$\langle\bullet\bullet\overline{p}\rangle$ are plotted with their
theoretical counterpart.

Now, $f(p)$ and $g(p)$ being known, the percolation probability can
be easily computed by using Eq.~(\ref{rel1}):
\begin{eqnarray}
n(p)=&p^{-2}&\left[\frac{(x-\tau)(2+\tau-x)}{\tau}\right]^\beta
\label{nth}\\
&\times & \left[\frac{(x-1+\tau)(x+1+\tau)}
{1+\tau}\right]^{1-\beta}\nonumber.
\end{eqnarray}
In Fig.~2, we compare this result with the numerically extrapolated
stationary percolation probability, and to the MF result of
Eq.~(\ref{MF}).
  
Finally, in Fig.~3, in order to test the validity of our basic
approximation Eq.~(\ref{fcm}), we plot $f(x_1,x_2)$ (defined in
Eq.~(\ref{f12})) as a function of $\max(x_1,x_2)$. We find that this
scatter plot is reasonably aligned around an effective curve, and that
$f_{MF}(x_1,x_2)=\rho_{MF}(\max(x_1,x_2))$ appears to be a lower bound
for the actual $f(x_1,x_2)$.

\begin{figure}[hhh]  
\psfig{file=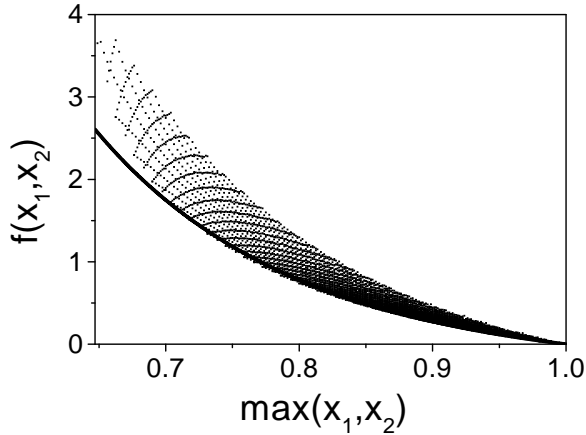,width=9cm}
\caption[]{We plot $f(x_1,x_2)$ defined in Eq.~(\ref{f12}) as a function of
$\max(x_1,x_2)$ (scatter plot). This is compared with the approximation
central to this paper: $f(x_1,x_2)=\rho_{MF}(\max(x_1,x_2))$ (thick
line). We observe that the theoretical expression seems to be a lower
bound for the actual $f(x_1,x_2)$, and that the dispersion (due to the
explicit dependence on $\min(x_1,x_2)$) is weak enough, so that the
scatter plot tends to align around an effective curve.}
\end{figure}

In conclusion, we have introduced a new approximation for a continuous
model equivalent to directed bond percolation. In this language, this
approximation amounts to properly modelizing the correlation function
$\rho_2(p_1,p_2)$, relating the properties of directed bond percolation
for two different percolation parameters $p_1$ and $p_2$. By assuming that
amplitude ratios are correctly described by mean field theory, we end
up with a precise description of the percolation probability. In
particular, we find an exponent $\beta$ in remarkable agreement with
the best available numerical simulations.

It would be interesting to exploit the present approach in order to
describe the dynamical properties of DP. This study is currently in
progress.

This approach could also prove useful in tackling the
notably difficult problem of parity conserving branching annihilating
walks \cite{dpreview}. This universality class is exemplified by the
reaction-diffusion model of diffusing particles $A$, involving
annihilation ($A+A\to\emptyset$) and branching ($A\to A+A+A$)
processes.  This problem has so far eluded all manner of theoretical
approaches in $d=1+1$.

\acknowledgments
I am very grateful to P.J. Basson for useful comments concerning this manuscript.


\begin{thebibliography}{99}
\bibitem{dp1} J. Marro and R. Dickman, {\it Nonequilibrium phase 
transitions in lattice models}, Cambridge University Press, Cambridge (1999).
\bibitem{dp2} W. Kinzel, in {\it Percolation Structures 
and Processes}, Ed. G. Deutscher, R. Zallen, and J. Adler,
Ann. Isr. Phys. Soc. {\bf 5}, 425 (Adam Hilgor, Bristol, 1983).
\bibitem{dpreview} H. Hinrichsen, Adv. Phys. {\bf 49}, 815 (2000).
\bibitem{dpapp} H. Hinrichsen, Braz. J. Phys. {\bf 30}, 69 (2000).
\bibitem{cata} R. M. Ziff, E. Gulari, and Y. Barshad, 
Phys. Rev. Lett. {\bf 56}, 2553 (1986).
\bibitem{surf} L.H. Tang and H. Leschhorn, 
Phys. Rev. A {\bf 45}, R8309 (1992).
\bibitem{sand} H. Hinrichsen, A. Jimenez-Dalmaroni, Y. Rozov, and E. Domany,
J. Stat. Phys. {\bf 98}, 1149 (2000).
\bibitem{dppomeau} Y. Pomeau, Physica D {\bf 23}, 3 (1986).
\bibitem{gen} P. Grassberger, {\it Directed percolation: 
results and open problems}, preprint WUB 96-2 (1996), unpublished.
\bibitem{pc} I. Jensen, Phys. Rev. Lett. {\bf 77}, 4988 (1996).
\bibitem{beta} I. Jensen,  J. Phys. A {\bf 32}, 5233 (1999).
\bibitem{cont1} A. Hansen and S. Roux, J. Phys. A {\bf 20}, L873 (1987).
\bibitem{cont2} P. Grassberger and Y.-C. Zhang, Physica A 
{\bf 224}, 169 (1996).
\bibitem{cont3} S. Maslov and Y.-C. Zhang, Physica A {\bf 223}, 1 (1996).
\bibitem{cont4} R. Dickman, M. A. Mu\~noz, A. Vespignani, and S. Zapperi, Braz.
J. Phys. {\bf 30}, 27 (2000).
\end{thebibliography}
\end{document}